\documentclass[11pt,a4paper,reqno,intlimits]{amsart}

\usepackage{amssymb}
\usepackage{chicago}
\usepackage{color}
\usepackage[mathscr]{euscript}
\usepackage{xspace}
\usepackage{enumerate}
\usepackage{epsfig,rotating}

\usepackage[bookmarksopen,pdfstartview=FitH]{hyperref}
\hypersetup{colorlinks,%
           citecolor=black,%
           filecolor=black,%
           linkcolor=black,%
           urlcolor=black}%

\begin{document}

\theoremstyle{plain}
\newtheorem{theorem}{Theorem}[section]
\newtheorem{lemma}[theorem]{Lemma}
\newtheorem{proposition}[theorem]{Proposition}
\newtheorem{claim}[theorem]{Claim}
\newtheorem{corollary}[theorem]{Corollary}
\newtheorem{axiom}{requirement}

\theoremstyle{definition}
\newtheorem{remark}[theorem]{Remark}
\newtheorem{note}{Note}[section]
\newtheorem{definition}[theorem]{Definition}
\newtheorem{example}[theorem]{Example}
\newtheorem*{ackn}{Acknowledgements}
\newtheorem{assumption}{Assumption}
\newtheorem{approach}{Approach}
\newtheorem{critique}{Critique}
\newtheorem{question}{Question}
\newtheorem{aim}{Aim}
\newtheorem*{asa}{Assumption ($\mathbf{A}$)}
\newtheorem*{asp}{Assumption ($\mathbb{P}$)}
\newtheorem*{ass}{Assumption ($\mathbb{S}$)}
\renewcommand{\theequation}{\thesection.\arabic{equation}}
\numberwithin{equation}{section}

\newcommand{\Law}{\ensuremath{\mathop{\mathrm{Law}}}}
\newcommand{\loc}{{\mathrm{loc}}}

\let\SETMINUS\setminus
\renewcommand{\setminus}{\backslash}

\def\stackrelboth#1#2#3{\mathrel{\mathop{#2}\limits^{#1}_{#3}}}

\newcommand\llambda{{\mathchoice
      {\lambda\mkern-4.5mu{\raisebox{.4ex}{\scriptsize$\backslash$}}}
      {\lambda\mkern-4.83mu{\raisebox{.4ex}{\scriptsize$\backslash$}}}
      {\lambda\mkern-4.5mu{\raisebox{.2ex}{\footnotesize$\scriptscriptstyle\backslash$}}}
      {\lambda\mkern-5.0mu{\raisebox{.2ex}{\tiny$\scriptscriptstyle\backslash$}}}}}

\newcommand{\prozess}[1][L]{{\ensuremath{#1=(#1_t)_{0\le t\le T}}}\xspace}
\newcommand{\prazess}[1][L]{{\ensuremath{#1=(#1_t)_{t\ge0}}}\xspace}
\newcommand{\pt}[1][N]{\ensuremath{\P_{T_{#1}}}\xspace}
\newcommand{\tk}[1][N]{\ensuremath{T_{#1}}\xspace}
\newcommand{\dd}[1][]{\ensuremath{\ud{#1}}\xspace}
\newcommand{\ttk}[1][k]{\ensuremath{[0,T_{#1}]}\xspace}

\newcommand{\scal}[2]{\ensuremath{\langle #1, #2 \rangle}}
\newcommand{\set}[1]{\ensuremath{\left\{#1\right\}}}
\newcommand{\ts}[1][]{\ensuremath{T_{#1}}\xspace}

\def\lev{L\'{e}vy\xspace}
\def\lk{L\'{e}vy--Khintchine\xspace}
\def\lib{LIBOR\xspace}
\def\mg{martingale\xspace}
\def\smmg{semimartingale\xspace}
\def\smmgs{semimartingales\xspace}

\def\lmm{LIBOR market model\xspace}
\def\fpm{forward price model\xspace}
\def\mfm{Markov-functional model\xspace}
\def\alm{affine LIBOR model\xspace}
\def\lmms{LIBOR market models\xspace}
\def\fpms{forward price models\xspace}
\def\mfms{Markov-functional models\xspace}
\def\alms{affine LIBOR models\xspace}

\def\half{\frac{1}{2}}

\def\F{\ensuremath{\mathcal{F}}}
\def\bF{\mathbf{F}}
\def\R{\ensuremath{\mathbb{R}}}
\def\Rp{\mathbb{R}_{\geqslant0}}
\def\Rm{\mathbb{R}_{\leqslant 0}}
\def\C{\ensuremath{\mathbb{C}}}
\def\U{\ensuremath{\mathcal{U}}}
\def\I{\mathcal{I}}
\def\K{\mathbb{K}}
\def\bk{\overline{K}}

\def\P{\ensuremath{\mathrm{I\kern-.2em P}}}
\def\E{\ensuremath{\mathrm{I\kern-.2em E}}}
\def\N{\ensuremath{\mathrm{I\kern-.2em N}}}

\def\ott{{0\leq t\leq T}}
\def\idd{{1\le i\le d}}

\def\icc{\mathpzc{i}}
\def\ecc{\mathbf{e}_\mathpzc{i}}

\def\uk{u_{k+1}}

\def\e{\mathrm{e}}
\def\ud{\ensuremath{\mathrm{d}}}
\def\dt{\ud t}
\def\ds{\ud s}
\def\dx{\ud x}
\def\dy{\ud y}
\def\dv{\ud v}
\def\dsdx{\ensuremath{(\ud s, \ud x)}}
\def\dtdx{\ensuremath{(\ud t, \ud x)}}

\def\lsnc{\ensuremath{\mathrm{LSNC-}\chi^2}}
\def\nc{\ensuremath{\mathrm{NC-}\chi^2}}

\def\tte{(t,T_k)}
\def\eps{\epsilon}

\def\krpt{\shortciteN{KellerResselPapapantoleonTeichmann09}}

\title{Old and new approaches to LIBOR modeling}

\author{Antonis Papapantoleon}

\address{Institute of Mathematics, TU Berlin, Stra\ss e des 17. Juni 136,
         10623 Berlin, Germany \& Quantitative Products Laboratory,
         Deutsche Bank AG, Alexanderstr. 5, 10178 Berlin, Germany}
\email{papapan@math.tu-berlin.de}
\date{}
\maketitle \pagestyle{myheadings}
\frenchspacing

\setcounter{tocdepth}{1}

\begin{abstract}
In this article, we review the construction and properties of some popular
approaches to modeling LIBOR rates. We discuss the following frameworks:
classical \lmms, \fpms and \mfms. We close with the recently developed \alms.
\end{abstract}

\section{Introduction}
\label{intro}

Interest rate markets are a large and important part of global financial markets.
The figures published by the Bank for International Settlements (BIS) show that
interest rate derivatives represent more than 60\% of the over-the-counter
markets over the years, in terms of notional amount; cf. Table \ref{BIS}. Hence,
it is important to have models that can adequately describe the dynamics and
mechanics of interest rates.

There is a notable difference between interest rate markets and stock or foreign
exchange (FX) markets. While in the latter there is a single underlying to be
modeled, the stock price or the FX rate, in interest rate markets there is a
whole \emph{family} of underlyings to be modeled, indexed by the time of maturity.
This poses unique challenges for researchers in mathematical finance and has led
to some fascinating developments.

The initial approaches to interest rate modeling considered short rates or
instantaneous forward rates as modeling objects, and then deduced from them
tradable rates. More recently, \textit{effective rates}, i.e. tradable market
rates such as the LIBOR or swap rate, were modeled directly. Models for
effective rates consider only a discrete set of maturity dates, the so-called
\textit{tenor structure}, which consists of the dates when these rates are
fixed. A review of the different approaches to modeling interest rates is beyond
the scope of the present article. There are many excellent books available,
focusing on the theoretical and practical aspects of interest rate theory. We
refer the reader e.g. to \citeN{Bjoerk04}, \citeN{MusielaRutkowski97},
\citeN{Filipovic09}, or \citeN{BrigoMercurio06}.

The aim of this article is to review the construction and basic properties of
models for \lib rates. We consider the following popular approaches: \lmms,
\fpms and \mfms, as well as the recently developed class of \alms. In section
\ref{axioms} we will present and discuss some basic requirements that models
for LIBOR rates should satisfy. These are briefly: \textit{positivity} of \lib
rates, \textit{arbitrage freeness} and \textit{analytical tractability}.

There are two natural starting points for modeling LIBOR rates: the rate itself
and the forward price. Although they differ only by an additive and a
multiplicative constant, cf. \eqref{basic}, the model dynamics are noticeably
different, depending on whether the model is based on the LIBOR or the forward
price. In addition, the consequences from the point of view of econometrics are
also significant.

Modeling LIBOR rates directly, leads to positive rates and arbitrage-free
dynamics, but the model is not analytically tractable. On the other hand, models
for the forward price are analytically tractable, but \lib rates can become
negative. The only models that can respect all properties simultaneously are
\mfms and \alms.

The article is organized as follows: in section \ref{markets} we introduce some
basic notation for interest rates and in section \ref{axioms} we describe the
basic requirements for \lib models. In section \ref{lmm} we review the
construction of \lmms, describe its shortcomings and discuss some approximation
methods developed to overcome them. In section \ref{fpm} we review \fpms and in
section \ref{mfm} we discuss \mfms. Finally, in section \ref{alm} we present
\alms and in section \ref{extend} we outline the extensions of LIBOR models to
the multi-currency and default risk settings.

\begin{table}
 \begin{center}
  {\renewcommand{\arraystretch}{1.05}
  \begin{tabular}{lrrrr}
                       & Dec 2006  & Dec 2007 & Jun 2008 & Dec 2008\\
\hline
  Foreign exchange     &   40,271  &  56,238  &  62,983  &  49,753\\
  Interest rate        &  291,581  & 393,138  & 458,304  & 418,678\\
  Equity-linked        &    7,488  &   8,469  &  10,177  &   6,494\\
  Commodity            &    7,115  &   8,455  &  13,229  &   4,427\\
  Credit default swaps &   28,650  &  57,894  &  57,325  &  41,868\\
  Unallocated          &   43,026  &  71,146  &  81,708  &  70,742\\
\hline
  Total                &  418,131  & 595,341  & 683,726 & 591,963\\
   \end{tabular}}~\\[1ex]
  \caption{Amounts outstanding of over-the-counter (OTC) derivatives
           by risk category and instrument (in billions of US dollars).
           Source: BIS Quarterly Review, September 2009.}
  \label{BIS}
  \end{center}
\end{table}

\section{Interest rate markets -- notation}
\label{markets}

Let us consider a discrete tenor structure $0=T_0<T_1<\dots<T_N$, with constant
tenor length $\delta=\tk[k+1]-T_k$. The following notation is introduced for
convenience; $K:=\{1,\dots,N-1\}$ and $\bk:=\{1,\dots,N\}$. Let us denote
by $B(t,T)$ the time-$t$ price of a \textit{zero coupon bond} maturing at time
$T$; by $L(t,T)$ the time-$t$ \textit{forward \lib rate} settled at time $T$
and received at time $T+\delta$; and by $F(t,T,U)$ the time-$t$ \textit{forward
price} associated to the dates $T$ and $U$. These fundamental quantities are
related by the following basic equation:
\begin{align}\label{basic}
  1+\delta L(t,T) = \frac{B(t,T)}{B(t,T+\delta)} = F(t,T,T+\delta).
\end{align}

Throughout this work, $\mathscr{B}=(\Omega,\F,\bF,\P)$ denotes a complete
stochastic basis, where $\F=\F_T$, $\bF=(\F_t)_{t\in[0,T]}$, and
$T_N\le T<\infty$. We denote by $\mathcal{M}(\P)$ the class of martingales on
$\mathscr{B}$ with respect to the measure $\P$.

We associate to each date $T_k$ in the tenor structure a \emph{forward martingale
measure}, denoted by $\pt[k]$, $k\in\bk$. By the definition of forward measures,
cf. \citeN[Def. 14.1.1]{MusielaRutkowski97}, the bond price with maturity $T_k$
is the numeraire for the forward measure $\pt[k]$. Thus, we have that forward
measures are related to each other via
\begin{align}\label{Pk-to-next}
\frac{\dd \P_{\tk[k]}}{\dd \P_{\tk[k+1]}} \Big|_{\F_t}
  = \frac{F(t,\tk[k],\tk[k+1])}{F(0,\tk[k],\tk[k+1])}
  = \frac{B(0,\tk[k+1])}{B(0,\tk[k])}
    \times \frac{B(t,\tk[k])}{B(t,\tk[k+1])},
\end{align}
while they are related to the terminal forward measure via
\begin{align}\label{Pk-to-final}
\frac{\dd \P_{\tk[k]}}{\dd \P_{\tk[N]}} \Big|_{\F_t}
  = \frac{F(t,\tk[k],\tk[N])}{F(0,\tk[k],\tk[N])}
  = \frac{B(0,\tk[N])}{B(0,\tk[k])}
    \times \frac{B(t,\tk[k])}{B(t,\tk[N])}.
\end{align}
All forward measures are assumed to be equivalent to the measure $\P$.

\section{Axioms for \lib models}
\label{axioms}

In this section, we present and discuss certain requirements that a model for
\lib rates should satisfy. These requirements are motivated both by the economic
and financial aspects of \lib rates, as well as by the practical demands
for implementing and using a model in practice. The aim here is to unify the
line of thought in \shortciteN{HuntKennedyPelsser00} and in 
\shortciteANP{KellerResselPapapantoleonTeichmann09}
\citeyear{KellerResselPapapantoleonTeichmann09}.

A model for \lib rates should satisfy the following \emph{requirements}:
\begin{itemize}
\item[(\textbf{A1})] \lib rates should be \emph{non-negative}:
            $L(t,T)\ge0$ for all $t\in[0,T]$.
\item[(\textbf{A2})] The model should be \emph{arbitrage-free}:
            $L(\cdot,T)\in\mathcal{M}(\P_{T+\delta})$.
\item[(\textbf{A3})] The model should be \emph{analytically tractable},
            easy to implement and quick to calibrate to market data.
\item[(\textbf{A4})] The model should provide a \emph{good calibration} to
            market data of liquid derivatives, i.e. caps and swaptions.
\end{itemize}

Requirements (A1) and (A2) are logical conditions originating from economics and
mathematical finance. Below, we briefly elaborate on (A3) and (A4); they stem
from practical demands and are more difficult to quantify precisely. In order to
clarify their difference, we point out that e.g. the Black--Scholes model
obviously satisfies (A3) but not (A4).

Requirement (A3) means that we can price liquid derivatives, e.g. caps and
swaptions in ``closed form'' in the model, so that the model can be calibrated
to market data in a fast and easy manner. Ideally, of course we would like to
be able to price as many derivatives as possible in closed form. Here, ``closed
form'' is understood in a broad sense meaning e.g. Fourier transform methods;
really closed form solutions \'a la Black--Scholes are typically hard to achieve.
\shortciteN{HuntKennedyPelsser00} say that the model is analytically tractable
if it is driven by a \emph{low-dimensional} Markov process. In \krpt, as well as
in the present article, we say that a model is analytically tractable if the
structure of the driving process is \emph{preserved} under the different forward
measures.

Finally, requirement (A4) means that the model is able to describe the
observed data accurately, without overfitting them. We will not examine this
requirement further in this article. On an intuitive level, since the models we
will describe in the sequel are driven by general Markov processes or general
semimartingales, we can always find a driving process that provides a good
calibration to market data. However, an empirical analysis should be performed
in order to identify such a driving process (cf. e.g. \shortciteNP{JarrowLiZhao07}
and \citeNP[Ch. III]{Skovmand08}).

\section{LIBOR market models}
\label{lmm}

\lmms were introduced in the seminal papers of 
Miltersen et al. \citeyear{MiltersenSandmannSondermann97},
\shortciteN{BraceGatarekMusiela97}
and \citeN{Jamshidian97}. In this framework, \lib rates are modeled as the exponential
of a Brownian motion under their corresponding forward measure, hence they are
log-normally distributed. This is the so-called \emph{log-normal LIBOR market model}.
Caplets are then priced by Black's formula (cf. \citeNP{Black76}), which is in
accordance with standard market practice. Later, \lmms were extended to
accommodate more general driving processes such as \lev processes, stochastic volatility
processes and general semimartingales, in order to describe more accurately the market
data; cf. \citeN{Jamshidian99}, Glasserman and Kou \citeyear{GlassermanKou03},
\citeN{EberleinOezkan05}, \citeANP{AndersenBrothertonRatcliffe05}
\citeyear{AndersenBrothertonRatcliffe05}, \citeN{BelomestnySchoenmakers06} and
\shortciteN{BelomestnyMathewSchoenmakers06}, to mention just a fraction of the existing
literature.

Consider an initial tenor structure of non-negative \lib rates $L(0,T_k)$, $k\in\bk$,
and let $\lambda(\cdot,T_k):[0,T_k]\to\R$ denote the volatility of the forward \lib
rate $L(\cdot,T_k)$, $k\in K$; the volatilities are assumed deterministic, for
simplicity. Let $H$ denote a semimartingale on $(\Omega,\F,\bF,\pt)$ with triplet
of semimartingale characteristics $\mathbb{T}(H|\pt)=(B,C,\nu)$ and $H_0=0$ a.s.;
$H$ satisfies certain integrability assumptions which are suppressed for brevity
(e.g. finite exponential moments, absolutely continuous characteristics). The
process $H$ is driving the dynamics of \lib rates and is chosen to have a
\emph{tractable} structure under $\pt$ (e.g. $H$ is a \lev or an affine process).

In \lmms, forward \lib rates are modeled as follows:
\begin{align}\label{LMM-1}
L(t,T_k) &= L(0,T_k)
 \exp\left( \int_0^t \beta(s,T_k)\ds + \int_0^t \lambda(s,T_k)\ud H_s \right),
\end{align}
where $\beta(\cdot,T_k)$ is the drift term that makes
$L(\cdot,T_k)\in\mathcal{M}(\pt[k+1])$, for all $k\in K$. Therefore, the model
clearly satisfies requirements (A1) and (A2).

Now, using Theorem 2.18 in \citeN{KallsenShiryaev02}, we have that
\begin{align}\label{LMM-2}
\beta(s,T_k)
 &= -\lambda(s,T_k)b_s^{\tk[k+1]}
    -\frac12\lambda^2(s,T_k)c_s \nonumber\\
 &\quad
    -\int_{\R}\big( \e^{\lambda(s,T_k)x}
                     - 1 - \lambda(s,T_k)x \big) F_s^{\tk[k+1]}(\dx),
\end{align}
such that indeed $L(\cdot,T_k)\in\mathcal{M}(\pt[k+1])$. Here
$(b_s^{\tk[k+1]}, c_s, F_s^{\tk[k+1]})$ denote the differential characteristics
of $H$ under $\pt[k+1]$.
Therefore, in order to completely understand the dynamics of the model we have
to calculate the characteristics $(b_s^{\tk[k+1]}, c_s, F_s^{\tk[k+1]})$.

These characteristics follow readily from Girsanov's theorem for semimartingales
(cf. \citeNP[III.3.24]{JacodShiryaev03}) once we have the density between the
measure changes at hand. It is convenient to express this density as a stochastic
exponential. Keeping \eqref{Pk-to-final} in mind, and denoting \eqref{LMM-1} as
follows
\begin{align}\label{LMM-3}
\ud L(t,T_k) = L(t-,T_k)\ud \widetilde{H}_t^k,
\end{align}
i.e. $\widetilde{H}^k$ is the exponential transform of the exponent in
\eqref{LMM-1}, we get from \eqref{basic} that
\begin{align}\label{LIBOR-3}
\ud F(t,T_k,T_{k+1})
 &= \delta\, \ud L(t,T_k) \nonumber\\
 &= \delta L(t-,T_k)\ud \widetilde{H}_t^k \nonumber\\
 &= F(t-,T_k,T_{k+1}) \frac{\delta L(t-,T_k)}{1+\delta L(t-,T_k)}
    \ud \widetilde{H}_t^k \nonumber\\
 \Rightarrow
F(t,T_k,\tk[k+1])
 &= F(0,T_k,\tk[k+1])\,
    \mathcal{E}\!\left( \int_0^\cdot\frac{\delta L(s-,T_k)}{1+\delta L(s-,T_k)}
                                  \ud \widetilde{H}_s^k \right)_{\!t}.
\end{align}
Therefore, the density between the measure changes takes the form
\begin{align}\label{LMM-4}
\frac{\dd \P_{\tk[k+1]}}{\dd \P_{\tk[N]}} \Big|_{\F_t}
  = \frac{B(0,\tk[N])}{B(0,\tk[k+1])}
    \times \prod_{l=k+1}^{N-1}
     \mathcal{E}\!\left( \int_0^\cdot
       \frac{\delta L(s-,T_l)}{1+\delta L(s-,T_l)}\ud \widetilde{H}_s^l \right)_{\!t}.
\end{align}

This calculation reveals the problem of \lmms: the density process between the
measure changes -- and thus the characteristics of $H$ under the forward measures
-- does not depend \emph{only} on the dynamics of $\widetilde{H}^k$, or
equivalently on the dynamics of $\int\lambda(s,T_k)\ud H_s$, as is the case in e.g.
HJM models. It also crucially depends on \emph{all subsequent} \lib rates, as the
product and the terms $\frac{\delta L(\cdot,T_l)}{1+\delta L(\cdot,T_l)}$ in
\eqref{LMM-4} clearly indicate. This means, in particular, that the structure of
the model is \emph{not preserved} under the different forward measures; e.g. if
$H$ is a \lev or an affine process under the terminal measure $\pt$, then $H$ is
neither a \lev nor an affine process under any other forward measure -- not even
a time-inhomogeneous version of those. Therefore, \lmms do not satisfy
requirement (A3).

The semimartingale $H$, that drives the dynamics of LIBOR rates, has the following
canonical decomposition under the terminal martingale measure $\pt$
\begin{align}\label{LMM-4.5}
 H_t = B_t + \int_0^t \sqrt{c_s}\ud W_s + \int_0^t\int_\R x(\mu^H-\nu)\dsdx,
\end{align}
(cf. \citeNP[II.2.38]{JacodShiryaev03} and \citeNP[Theorem 3.4.2]{KaratzasShreve91})
where $W$ denotes the $\pt$-Brownian motion and $\mu^H$ denotes the random measure
of the jumps of $H$. The $\pt$-compensator of $\mu^H$ is $\nu$ and
$C=\int_0^\cdot c_s\ds$. Straightforward calculations using the density in
\eqref{LMM-4} \linebreak (cf. e.g. \citeNP{Kluge05} or \citeNP{PapapantoleonSiopacha09})
yield that the $\pt[k+1]$-Brownian motion $W^{T_{k+1}}$ is related to the
$\P_{T_N}$-Brownian motion via
\begin{align}\label{LMM-5}
W_t^{T_{k+1}}
 & = W_t - \int_0^t \left(\sum_{l=k+1}^{N-1}
            \frac{\delta L(t-,T_l)}
                 {1+\delta L(t-,T_l)}\lambda(t,T_l)\right)\sqrt{c_s} \ud s,
\end{align}
while the $\pt[k+1]$-compensator of $\mu^H$, $\nu^{T_{k+1}}$, is related to the
$\P_{T_N}$-compensator of $\mu^H$ via
\begin{align}\label{LMM-6}
\nu^{T_{k+1}}\dsdx
 &= \left(\prod_{l=k+1}^{N-1}\gamma(s,x,T_l)\right)\nu\dsdx,
\end{align}
where
\begin{align}\label{LMM-7}
\gamma(s,x,T_l,)
 = \frac{\delta L(s-,T_l)}{1+\delta L(s-,T_l)}\Big(\e^{\lambda(s,T_l)x}-1\Big) +1.
\end{align}
In addition, the drift term of the \lib rate $L(\cdot,T_k)$ relative to the
$\pt$ differential characteristics of $H$, i.e. $(b,c,F)$, is
\begin{align}\label{LMM-8}
\widehat{\beta}(s,\ts[k])
 & = -\frac12 \lambda^2(s,T_k) c_s
     - c_s \lambda(s,T_k)
        \sum_{l=k+1}^{N-1}\frac{\delta L(s-,T_l)}{1+\delta L(s-,T_l)}\lambda(s,T_l)
  \nonumber\\ &\,\,\,\,
     - \int_{\R}\left(\Big(\e^{\lambda(s,T_k) x}-1\Big)
          \prod_{l=k+1}^{N-1}\gamma(s,x,T_l) - \lambda(s,T_k) x\right)F_s(\ud x).
\end{align}

The consequences of the intractability of \lmms are the following. When the
driving process is a \emph{continuous} semimartingale, then
\begin{itemize}
\item caplets can be priced in closed form;
\item swaptions and other multi-LIBOR products cannot be priced in closed form;
\item Monte-Carlo simulations are particularly time consuming, since one is
      dealing with coupled high dimensional stochastic processes.
\end{itemize}
When the driving process is a \emph{general} semimartingale, then
\begin{itemize}
\item even caplets cannot be priced in closed form, let alone swaptions
      or other multi-\lib derivatives;
\item the Monte-Carlo simulations are equally time consuming.
\end{itemize}

Several approximation methods have been developed in the literature in order to
overcome these problems. We briefly review three of the proposed methods below;
for other methods and empirical comparison we refer the interested reader to
the review paper by \citeN{JoshiStacey08}.

\subsection{``Frozen drift'' approximation}
The first and easiest solution to the problem is the so-called ``frozen drift''
approximation, where one replaces the random terms in \eqref{LMM-8} or
\eqref{LMM-4} by their deterministic initial values, i.e.
\begin{align}\label{LMM-FD1}
\frac{\delta L(s-,T_l)}{1+\delta L(s-,T_l)}
 \approx \frac{\delta L(0,T_l)}{1+\delta L(0,T_l)}.
\end{align}
This approximation was first proposed by \shortciteN{BraceGatarekMusiela97}, and
has been used by many authors ever since. Under this approximation, the
measure change becomes a structure preserving one -- observe that the density in
\eqref{LMM-4} depends now only on the driving process and the volatility
structure -- and the resulting \emph{approximate} \lmm is analytically tractable;
e.g. caplets and swaptions can now be priced in closed form even in models
driven by semimartingales with jumps.

\begin{figure}
 \centering
  \includegraphics[width=177pt]{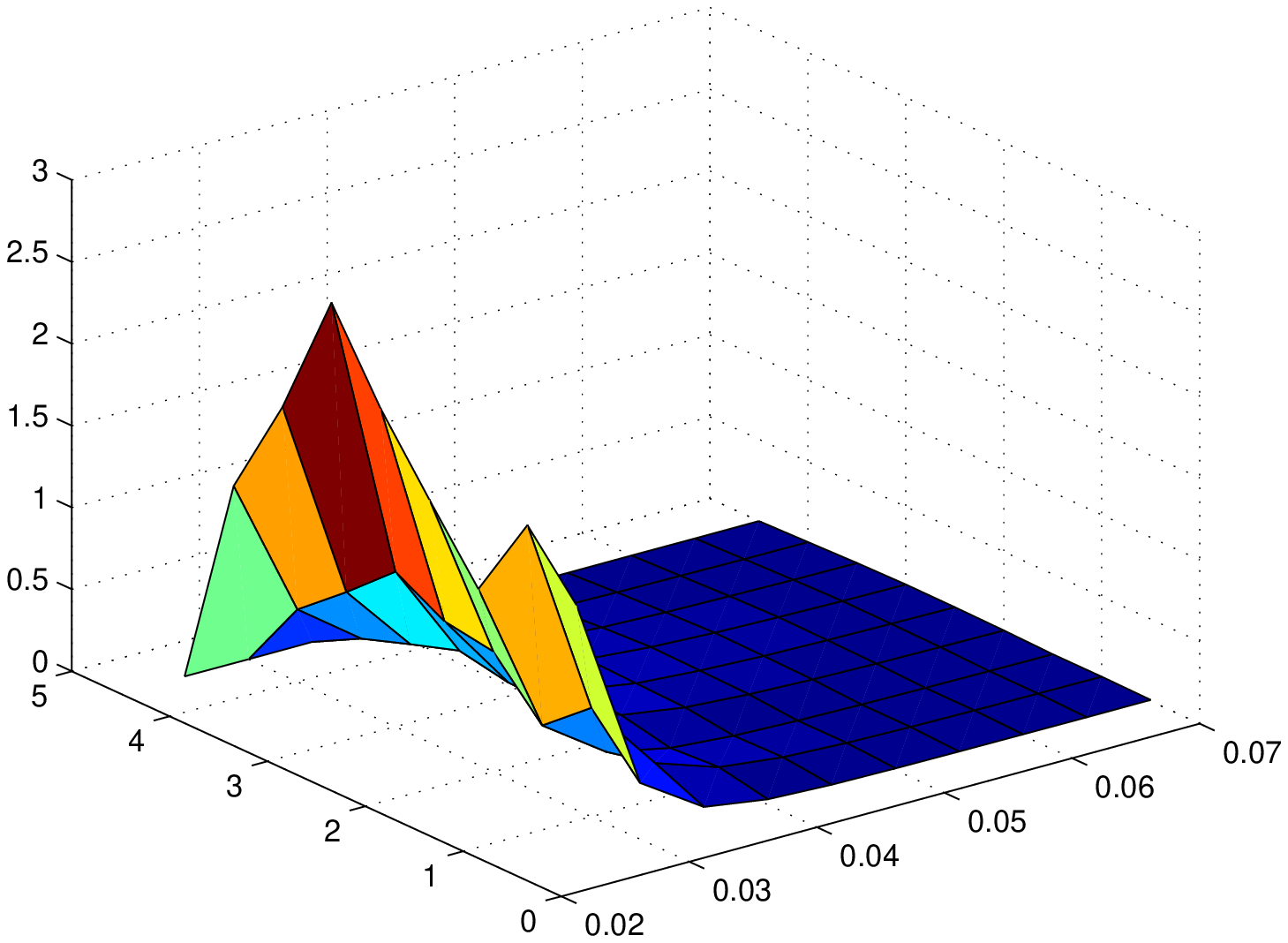}
  \includegraphics[width=177pt]{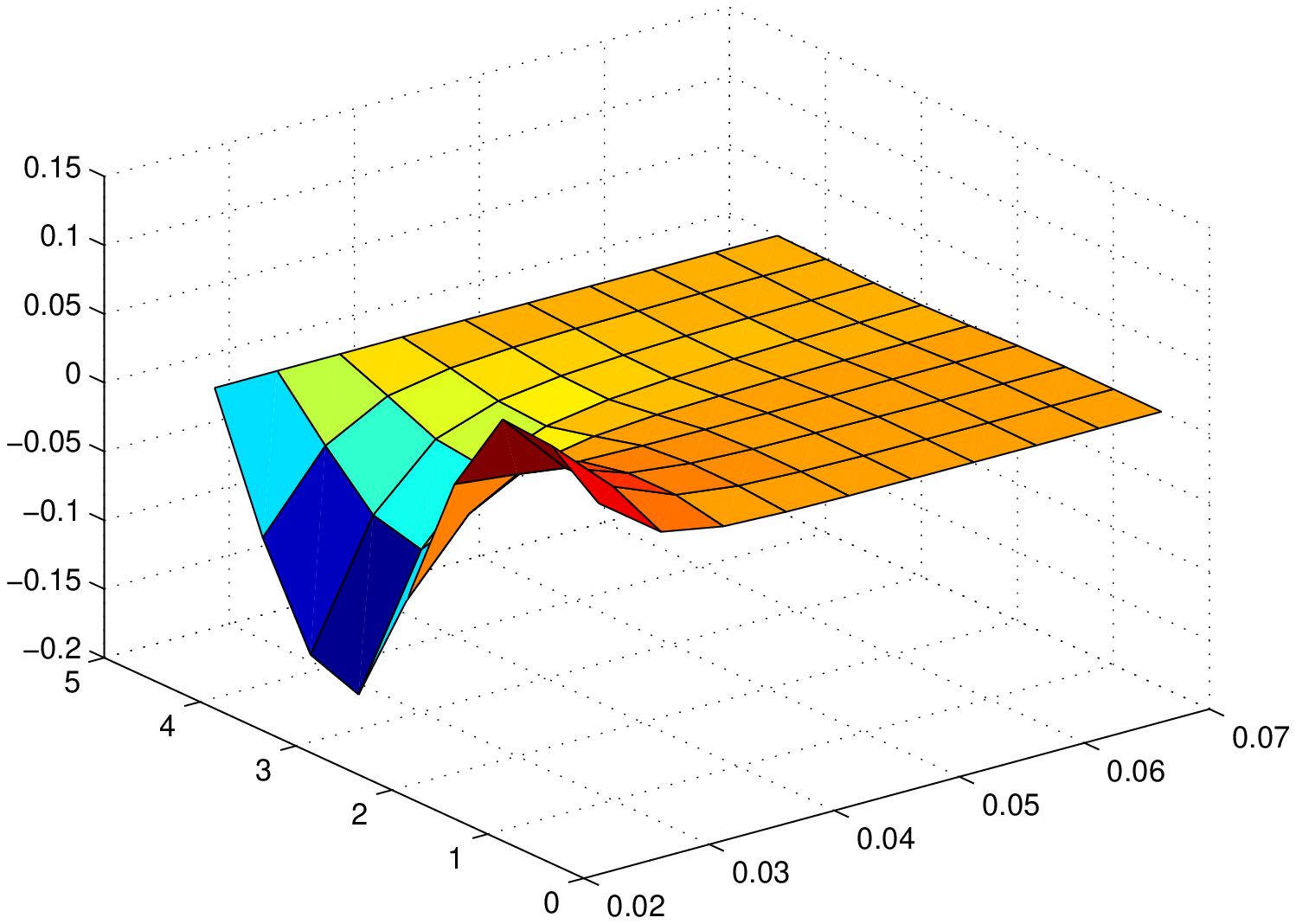}
 \caption{Difference in implied volatilities between the actual LIBOR and the
          frozen drift prices (left), and between the actual LIBOR and the
          Taylor approximation prices (right). Source: Papapantoleon
          and Siopacha (2009).}
 \label{diffs}
\end{figure}

However, this method yields poor empirical results. Comparing the prices of
either liquid options, or long-dated options, using the frozen drift approximation
to the prices obtained by the simulation of the actual dynamics for the LIBOR rates,
we can observe notable differences both in terms of prices and in terms of
implied volatilities. See Figure \ref{diffs} for an example. We refer to
\shortciteN{KurbanmuradovSabelfeldSchoenmakers}, \citeN{SiopachaTeichmann07},
and \citeN{PapapantoleonSiopacha09} for further numerical illustrations.

\subsection{Log-normal approximations}
The following approximation schemes for the log-normal \lmm were developed by
Kurbanmuradov et al.
\citeyear{KurbanmuradovSabelfeldSchoenmakers}. Consider the log-normal \lmm
driven by a one-dimensional Brownian motion, for simplicity. The dynamics of
LIBOR rates (expressed under the terminal measure) take the form
\begin{align}\label{LMM-LG-1}
\ud L(t,T_k)
 = L(t,T_k)\big( \lambda_t(T_k)\ud W_t + \beta_t(T_k)\dt \big),
\end{align}
where the drift term equals
\begin{align}\label{LMM-LG-1+1/2}
\beta_t(T_k)
 &= -\lambda_t(T_k) \sum_{l=k+1}^{N-1}
     \frac{\delta L(t-,T_l)}{1+\delta L(t-,T_l)}\lambda_t(T_l),
\end{align}
cf. \eqref{LMM-8}; w.l.o.g. we can set $c\equiv1$. A very crude log-normal
approximation is to ``neglect'' the non-Gaussian terms in the SDE, i.e. to set
$\beta_t(\ts[k])\equiv0$. Of course, this approximation does not yield
satisfactory results -- in principle, results are even worse than the frozen
drift approximation.

One can develop more refined log-normal approximations as follows: let
$f(x)=\frac{\delta x}{1+\delta x}$, and define the process $Z$, which equals the
terms that need to be approximated; i.e.
\begin{align}\label{LMM-LG-2}
Z(t,T_k)
 = \frac{\delta L(t,T_k) }{1+\delta L(t,T_k)}
 =f\big( L(t,T_k) \big).
\end{align}
Applying It\^o's formula, we derive the SDE that $Z(\cdot,T_k)$ satisfies
\begin{align}\label{LMM-LG-3}
\ud Z(t,T_k)
 &= f'(L(t,T_k))L(t,T_k)\lambda_t(T_k)\ud W_t \nonumber\\
 &\,\,+ \big\{f'(L(t,T_k))L(t,T_k)\beta_t(T_k)
  + \frac12 f''(L(t,T_k))L^2(t,T_k)\lambda^2_t(T_k)\big\}\dt \nonumber\\
 &=: A_k(t,Z)\dt + B_k(t,Z)\ud W_t,
\end{align}
with the initial condition $Z(0,T_k)=f(L(0,T_k))$. Note that the coefficients
$A_k$ and $B_k$ can be calculated explicitly, by solving \eqref{LMM-LG-2} for
$L$ and substituting into \eqref{LMM-LG-3}. Moreover, $Z$ in $A_k$ and $B_k$ denotes
the dependence on the whole vector $Z=[Z(\cdot,T_1),\dots,Z(\cdot,T_N)]$. The
first and second \emph{Picard iterations} for the solution of this SDE are
\begin{align}\label{LMM-LG-4}
Z^0(t,T_k)
 &= Z(0,T_k)
  = \frac{\delta L(0,T_k) }{1+\delta L(0,T_k)},
\end{align}
and
\begin{align}\label{LMM-LG-5}
Z^1(t,T_k)
 &= Z(0,T_k) + \int_0^t A_k(s,Z^0)\ds + \int_0^t B_k(s,Z^0)\ud W_s.
\end{align}
Note that $Z^0$ is constant, while $Z^1(t,T_k)$ has a Gaussian distribution
since the coefficients $A_k(\cdot,Z^0)$ and $B_k(\cdot,Z^0)$ are deterministic.

Now, replacing the random terms $Z(\cdot,T_k)$ in $\beta(T_k)$ with the Picard iterates
$Z^0(\cdot,T_k)$ and $Z^1(\cdot,T_k)$ leads to two different \emph{log-normal
approximations} to the dynamics of \lib rates. Obviously, the approximation by
$Z^0$ is nothing else than the frozen drift approximation. The approximation by
$Z^1$ is again log-normal, cf. \eqref{LMM-LG-1+1/2} and \eqref{LMM-LG-2}, and
yields very good empirical results. This latter approximation has the advantage
that the law of the approximate \lib rate is known at any time $t$, hence the
time-consuming Monte Carlo simulations can be avoided. For the empirical and
numerical analysis of these approximations we refer to
\shortciteN{KurbanmuradovSabelfeldSchoenmakers} and \citeN[Ch. 6]{Schoenmakers05}.

\subsection{Strong Taylor approximation}
Another approximation method has been recently developed by
\citeN{SiopachaTeichmann07} and Papapantoleon and Siopacha
\citeyear{PapapantoleonSiopacha09}. The main idea
behind this method is to replace the random terms in the drift \eqref{LMM-8} by
their first-order strong Taylor approximations. The Taylor approximation is developed
by a perturbation of the SDE for the \lib rates and a subsequent Taylor expansion.

Let us denote the log-\lib rates by $G(\cdot,T_k):=\log L(\cdot,T_k)$. Then, they
satisfy the linear SDE (under the terminal measure)
\begin{align}\label{LMM-ST-1}
\ud G(t,T_k) = \widehat{\beta}(t,T_k)\dt + \lambda\tte\ud H_t,
\end{align}
with initial condition $G(0,T_k)=\log L(0,T_k)$; cf. \eqref{LMM-1} and \eqref{LMM-8}.
We perturb this SDE by a real parameter $\eps$, i.e.
\begin{align}\label{LMM-ST-2}
\ud G^\eps(t,T_k) = \eps\big(\widehat{\beta}^\eps(t,T_k)\dt + \lambda\tte\ud H_t\big),
\end{align}
where $G^\eps(0,T_k)=G(0,T_k)$ for all $\eps$. The superscript $\eps$ in the
drift term $\widehat{\beta}^\eps(\cdot,T_k)$ is a reminder that this term is
also perturbed by $\eps$, since it contains all subsequent \lib rates; see
\eqref{LMM-8} again. Now, the \emph{first order strong Taylor approximation}
of $G^\eps$, denoted by $\mathbf{T}G^\eps$, is
\begin{align}\label{LMM-ST-3}
\mathbf{T}G^\eps(t,T_k)
 = G(0,T_k)
  + \eps \frac{\partial}{\partial\eps}\big|_{\eps=0} G^\eps(t,T_k).
\end{align}
We denote the ``first variation'' process
$\frac{\partial}{\partial\eps}\big|_{\eps=0} G^\eps(\cdot,T_k)$ by $Y(\cdot,T_k)$,
and then we can deduce, after some calculations, that it has the decomposition
\begin{align}\label{LMM-ST-4}
Y(t,T_k)
 &= \int_0^t \widehat{\beta}^0(s,T_k)\ds + \int_0^t \lambda(s,T_k)\ud H_s,
\end{align}
where $\widehat{\beta}^0(s,T_k):=\widehat{\beta}^\eps(s,T_k)|_{\eps=0}$. Hence,
this is a \emph{deterministic} drift term, obtained by replacing the random
terms in \eqref{LMM-8} by their deterministic initial values. In particular, we
can easily deduce from \eqref{LMM-ST-4} that, for example, if $H$ is a \lev
process then $Y(\cdot,T_k)$ is a \emph{time-inhomogeneous} \lev process.

Concluding, we have developed the following approximation scheme:
\begin{align}\label{LMM-ST-5}
\log L(t,T_k) \approx \log L(0,T_k) + Y(t,T_k),
\end{align}
where $Y(\cdot,T_k)$ has the decomposition \eqref{LMM-ST-4}; compare with
\eqref{LMM-1}.

The advantage of this method is threefold: (a) it is universal, and can be applied
to \lib models with stochastic volatility and/or jumps, (b) it is faster and easier
to simulate than the actual SDE for the \lib rates, and (c) the empirical performance
is very satisfactory; cf. Figure \ref{diffs} and the aforementioned articles for
further numerical illustrations. The drawback is that it is based on Monte Carlo
simulations, hence computational times can become long.

\section{Forward price models}
\label{fpm}

Forward price models were developed by \citeN{EberleinOezkan05}, and further
investigated by \citeN{Kluge05}; see also \citeN{EberleinKluge06}.\!
We consider a setting similar to \lmms, i.e. an
initial tenor structure of non-negative \lib rates, $\lambda(\cdot,T_k)$
denotes the volatility of the forward \lib rate $L(\cdot,T_k)$, and $H$ denotes
a semimartingale on $(\Omega,\F,\bF,\pt)$ with triplet of characteristics
$(B,C,\nu)$; again some assumptions on $H$ are suppressed. The process $H$ is
driving the dynamics of \lib rates and has a \emph{tractable} structure under
$\pt$ (e.g. $H$ is a \lev or an affine process).

Instead of modeling the forward \lib rate directly, one now models the forward
price in a similar fashion; that is
\begin{align}\label{fp-model-1}
1\!+\!\delta L(t,T_k)
 &= (1\!+\!\delta L(0,T_k))
    \exp\left( \int_0^t \beta(s,T_k)\ds + \int_0^t\lambda(s,T_k)\ud H_s \right),
\end{align}
where again the drift term is such that $L(\cdot,T_k)\in\mathcal{M}(\pt[k+1])$,
for all $k\in K$; i.e. $\beta(\cdot,T_k)$ has similar form to \eqref{LMM-2}.
Therefore the model obviously satisfies requirement (A2).

Now, from \eqref{Pk-to-final} and \eqref{fp-model-1}, we get that the density
between the forward measures is
\begin{align}\label{fp-model-2}
\frac{\dd \P_{\tk[k+1]}}{\dd \P_{\tk[N]}} \Big|_{\F_t}
 &= \frac{B(0,\tk[N])}{B(0,\tk[k+1])}
    \times \prod_{l=k+1}^{N-1}\big(1+\delta L(t,T_l)\big)\\
 &= \frac{B(0,\tk[N])}{B(0,\tk[k+1])}
    \times \exp\left( \int_0^t\!\sum_{l=k+1}^{N-1} \beta(s,T_l)\ds +
     \int_0^t\!\sum_{l=k+1}^{N-1} \lambda(s,T_l)\ud H_s \right). \nonumber
\end{align}
Observe that this density only depends on the driving process $H$ and the volatility
structures, hence we can deduce that the measure changes between forward
measures are \emph{Esscher transformations}; cf. \citeANP{KallsenShiryaev02} 
\linebreak\citeyear{KallsenShiryaev02} for
the Esscher transform. Analogously to eqs.
\eqref{LMM-4.5}--\eqref{LMM-6}, we have now that the $\pt[k+1]$-Brownian motion
is related to the $\P_{T_N}$-Brownian motion via
\begin{align}\label{fp-model-3}
W_t^{T_{k+1}}
 & = W_t
   - \int_0^t \left(\sum_{l=k+1}^{N-1}\lambda(t,T_l)\right)\sqrt{c_s} \ud s,
\end{align}
while the $\pt[k+1]$-compensator of $\mu^H$, is related to the
$\P_{T_N}$-compensator of $\mu^H$ via
\begin{align}\label{fp-model-4}
\nu^{T_{k+1}}\dsdx
 &= \exp\left(x\sum_{l=k+1}^{N-1}\lambda(s,T_l)\right)\nu\dsdx.
\end{align}
Thus the structure of the driving process is preserved. For example, if $H$
is a \lev or an affine process under $\pt$, then it becomes a
\emph{time-inhomogeneous} \lev or affine process respectively under any forward
measure $\pt[k+1]$. This implies that requirement (A3) is satisfied, so that
caplets and swaptions can be priced in closed form. In this class of models, we
have the additional benefit that we can even price some exotic path-dependent
options easily using Fourier transform methods; see \citeN{KlugePapapantoleon06}
for an example.

The main shortcoming of \fpms is that \emph{negative} \lib rates can occur,
similarly to HJM models, since
\begin{align*}
1+\delta L(t,T_k)>0 \,\,\nRightarrow\,\, L(t,T_k)>0
\qquad\text{for all }\, t\in[0,T_k].
\end{align*}
Therefore, this model can violate requirement (A1).

\section{Markov-functional models}
\label{mfm}

\mfms were introduced in the seminal paper of Hunt, Kennedy, and Pelsser
\citeyear{HuntKennedyPelsser00}. In
contrast to the other approaches described in this review, the aim of \mfms is
not to model some fundamental quantity, e.g. \lib or swap rates, directly. Instead,
\mfms are constructed by inferring the model dynamics, as well as their functional
forms, through matching the model prices to the market prices of certain liquid
derivatives. That is, they are \emph{implied interest rate models}, and should be
thought of in a fashion similar to local volatility models and implied trees in
equity markets.

The main idea behind \mfms is that bond prices and the numeraire are, at any point
in time, a function of a \emph{low-dimensional} Markov process under some martingale
measure. The functional form for the bond prices is selected such that the model
accurately calibrates to the relevant market prices, while the freedom to choose the
Markov process makes the model realistic and tractable. Moreover, the functional form
for the numeraire can be used to reproduce the marginal laws of swap rates or other
relevant instruments for the calibration.

More specifically, let $(M,\mathbf{M})$ denote a numeraire pair, and consider a
(time-inhomogeneous) Markov process \prozess[X] under the measure $\mathbf{M}$.
In the framework of Markov-functional models, one assumes that bond prices have
the functional form
\begin{align}\label{MFM-1}
B(t,S) = \mathrm{B}(t,S;X_t),
 \quad 0\le t\le \partial_S\le S,
\end{align}
where $\partial_S$ denotes some ``boundary curve''. In applications, the boundary
curve typically  has the form
\begin{align}\label{MFM-2}
\partial_S
 = \left\{
     \begin{array}{ll}
       S,   & S\le T_{*}, \\
       T_*, & S>T_{*},
     \end{array}
   \right.
\end{align}
where $T_*$ is a common time of maturity. One further assumes that the numeraire
$M$ is also a function of the driving Markov process $X$, i.e.
\begin{align}\label{MFM-3}
M_t = \mathrm{M}(t;X_t),
 \quad 0\le t\le T.
\end{align}
Therefore, in order to specify a \mfm, the following quantities are required:
\begin{itemize}
\item[(P1)] the law of $X$ under the measure $\mathbf{M}$;
\item[(P2)] the functional form $\mathrm{B}(\partial_S,S;\cdot)$ for $S\in[0,T]$;
\item[(P3)] the functional form $\mathrm{M}(t;\cdot)$ for $0\le t\le T$.
\end{itemize}

In applications, the Markov process is specified first and is typically a diffusion
process with time-dependent volatility. Then, the functional forms for the bond
prices and the numeraire are implied by calibrating the model to market prices of
liquid options. The choice of the calibrating instruments depends on the exotic
derivative that should be priced or hedged with the model. If the exotic depends
on LIBOR rates, e.g. the flexible cap, then the model is calibrated to digital caplets,
which leads to the \emph{Markov-functional LIBOR model}. If the exotic depends on
swap rates, e.g. the Bermudan swaption, then the model is calibrated to digital swaptions,
which leads to the \emph{Markov-functional swap rate model}. Let us point out that the
functional forms are typically not known in closed form, and should be computed
numerically. These models typically satisfy requirements (A1), (A2) and (A3).
For further details and concrete applications we refer the reader to the books
by \citeN{HuntKennedy04} and \citeN{Fries07}, and the references therein.

\begin{remark}
Let us point out that \fpms and \alms, that will be introduced in section \ref{alm},
belong to the class of \mfms, while \lmms do not. In \lmms the \lib rates are
functions of a \textit{high-dimensional} Markov process. 
\end{remark}

\subsection{Markov-functional LIBOR model}
In order to gain a better understanding of the construction of \mfms, we will
briefly describe a \mfm calibrated to \lib rates. This model is called the
\textit{Markov-functional LIBOR model}.

The set of relevant market rates are \lib rates $L(\cdot,T_k), k\in K$. We will
consider the numeraire pair $(M,\mathbf{M})=(B(\cdot,T_N),\pt)$.

In order to be consistent with Black's formula for caplets, we assume that
$L(\cdot,T_{N-1})$ is a log-normal martingale under \pt, i.e.
\begin{align}\label{MFL-1}
\ud L(t,T_{N-1})
= \sigma(t,T_{N-1}) L(t,T_{N-1}) \ud W_t,
\end{align}
where $W$ denotes a standard Brownian motion under $\pt$ and $\sigma(\cdot,T_{N-1})$
is a deterministic, time-dependent volatility function. Hence, we have that
\begin{align}\label{MFL-2}
L(t,T_{N-1})
= L(0,T_{N-1}) \exp\Big( -\frac12\Sigma_t + X_t\Big),
\end{align}
where $\Sigma=\int_0^\cdot \sigma^2(s,T_{N-1})\ds$, and $X$ is a deterministic
time-change of the Brownian motion, that satisfies
\begin{align}\label{MFL-3}
\ud X_t = \sigma(t,T_{N-1})\ud W_t.
\end{align}
We will use $X$ as the driving process of the model, which specifies (P1).

Regarding (P2), the boundary curve is exactly of the form \eqref{MFM-2} with
$T_*=T_{N-1}$, hence we need to specify $\mathrm{B}(T_i,T_i;X_{T_i})$ for
$i\in K$, which is trivially the unit map. We also need to specify
$\mathrm{B}(T_{N-1},T_N;X_{T_{N-1}})$; using \eqref{basic} and
\eqref{MFL-2} we get that
\begin{align}\label{MFL-4}
\mathrm{B}(T_{N-1},T_N;X_{T_{N-1}})
= \frac{1}{1 + \delta L(0,T_{N-1}) \exp\big( -\frac12\Sigma_{T_{N-1}}+ X_{T_{N-1}}\big)}.
\end{align}
Then, we can recover bond prices in the interior of the region bounded by
$\partial_S$ using the martingale property:
\begin{align}\label{MFL-5}
\mathrm{B}(t,S;X_t)
= \mathrm{B}(t,T_N;X_t)
  \E_{T_N}\bigg[ \frac{\mathrm{B}(\partial_S,S;X_{\partial_S})}
                      {\mathrm{B}(\partial_S,T_N;X_{\partial_S})} \Big|\F_t \bigg].
\end{align}

Now, it remains to specify the functional form $\mathrm{B}(T_i,T_N;X_{T_i})$,
$i\in K$, for the numeraire, cf. (P3). In the framework of the Markov-functional
\lib model, this is done by deriving the numeraire from \lib rates and inferring
the functional forms of the \lib rates via calibration to market prices. Equation
\eqref{basic} combined with \eqref{MFM-2} and the fact that $B(T_i,T_{i+1})$ is
a function of $X_{T_i}$, cf. \eqref{MFL-5}, yield that $L(T_i,T_i)$ is also a
function of $X_{T_i}$. The functional form is
\begin{align}\label{MFL-6}
1+\delta\mathrm{L}(T_i,T_i;X_{T_i})
&= \frac{1}{\mathrm{B}(T_i,T_{i+1};X_{T_{i}})} \nonumber\\
&= \frac{1}{\mathrm{B}(T_i,T_N;X_{T_i})
  \E_{T_N}\Big[ \frac{1}{\mathrm{B}(T_{i+1},T_N;X_{T_{i+1}})} \Big|\F_{T_i} \Big]}.
\end{align}
Rearranging, we get the following functional form for the numeraire
\begin{align}\label{MFL-7}
\mathrm{B}(T_i,T_N;X_{T_i})
&= \frac{1}{(1+\delta\mathrm{L}(T_i,T_i;X_{T_i}))
   \E_{T_N}\Big[ \frac{1}{\mathrm{B}(T_{i+1},T_N;X_{T_{i+1}})} \Big|\F_{T_i} \Big]}.
\end{align}
This formula provides a backward induction scheme to calculate
$\mathrm{B}(T_i,T_N;\cdot)$ from $\mathrm{B}(T_{i+1},T_N;\cdot)$ for any value
of the Markov process; the induction starts from $B(T_N,T_N)=1$.

The calibrating instruments are digital caplets with payoff
$1_{\{L(T_i,T_i)>\mathcal{K}\}}$, $i\in K$, and their market values are provided
by Black's formula; we denote them by $\mathbb{V}_0(T_i,\mathcal{K})$.
Assuming that the map $\xi\mapsto\mathrm{L}(T_i,T_i;\xi)$ is strictly increasing,
there exists a unique strike $\mathcal{K}(T_i,x^*)$ such that the set equality
\begin{align}\label{MFL-10}
\big\{X_{T_i}>x^*\big\}
 = \big\{\mathrm{L}(T_i,T_i;X_{T_i})>\mathcal{K}(T_i,x^*)\big\}
\end{align}
holds almost surely. Define the model prices
\begin{align}\label{MFL-11}
\mathbb{U}_0(T_i,x^*)
 = B(0,T_{N})
   \E_{T_N}\bigg[\frac{\mathrm{B}(T_i,T_{i+1};X_{T_{i}})}
                     {\mathrm{B}(T_i,T_N;X_{T_{i}})} 1_{\{X_{T_i}>x^*\}}\bigg],
\end{align}
which have to be calculated numerically. Therefore, we can equate market and
model prices
\begin{align}\label{MFL-12}
\mathbb{V}_0(T_i,\mathcal{K}(T_i,x^*)) = \mathbb{U}_0(T_i,x^*),
\end{align}
where the strike $\mathcal{K}(T_i,x^*)$ is determined by Black's formula using
some numerical algorithm.

Hence, we have specified, numerically at least, the functional form for the
LIBOR rates, which yields also the functional form for the numeraire via \eqref{MFL-7}.
This completes the specification of the Markov-functional LIBOR model. This
model satisfies requirement (A3), in the sense of \shortciteANP{HuntKennedyPelsser00}
\citeyear{HuntKennedyPelsser00},
since all bond prices are functions of a one-dimensional diffusion.

\section{Affine LIBOR models}
\label{alm}

Affine \lib models were recently developed by Keller-Ressel, Papapantoleon, and Teichmann
\citeyear{KellerResselPapapantoleonTeichmann09} with the aim of combining the
advantages of \lmms and \fpms, while circumventing their drawbacks. We provide
here a more general outline of this framework, which is based on two key
ingredients: martingales \emph{greater than 1}, which are \emph{increasing} in
some parameter.

The construction of martingales greater than 1 is done as follows: let $Y_T^u$
be an $\F_T$-measurable, integrable random variable, taking values in $[1,\infty)$,
and set
\begin{align}\label{Mge1}
M_t^u = \E[Y_T^u|\F_t],
\qquad \ott.
\end{align}
Then, using the tower property of conditional expectations, it easily follows that
\prozess[M^u] is a martingale. Moreover, it obviously holds that $M_t^u\ge1$ for
all $t\in[0,T]$.

In addition, assume that the map $u\mapsto Y_T^u$ is \emph{increasing}; then, we
immediately get that the map
\begin{align}\label{Mincrease}
u\mapsto M_t^u
\end{align}
is also increasing, for all $t\in[0,T]$. 

Now, using the family of martingales $M^u$ we can model quotients of bond prices
as follows. Consider a \emph{decreasing} sequence $(u_k)_{k\in\bk}$ and set
\begin{align}\label{bond-quots}
\frac{B(t,T_k)}{B(t,T_N)}
 &= M_t^{u_k},
 \qquad t\in\ttk,\, k\in\bk,
\end{align}
requiring that the initial values of the martingales fit today's observed market
prices, i.e. $\frac{B(0,T_k)}{B(0,T_N)}= M_0^{u_k}$. Since $M^u$ is increasing
in $u$, we have that
\begin{align}\label{order-M}
M_t^{u_k} \ge M_t^{u_l}
 \quad\text{ for }\quad
k\le l \Leftrightarrow u_k\ge u_l.
\end{align}
Hence, we can deduce that bond prices are decreasing as functions of time of
maturity, i.e. $B(t,T_k)\ge B(t,T_l)$ for $k\le l$.

Turning our attention to LIBOR rates, we get that
\begin{align}\label{LIBOR-1}
1 + \delta L(t,T_k)
 &= \frac{B(t,T_k)}{B(t,\tk[k+1])}
  = \frac{M_t^{u_k}}{M_t^{\uk}}
 \ge 1,
\end{align}
for all $t\in\ttk$ and all $k\in K$; this is a trivial consequence of \eqref{order-M}.
Moreover, the martingale property of the \lib rate under its corresponding forward
measure follows easily from the structure of the measure changes \eqref{Pk-to-final},
and the structure of the martingales. Indeed, we have that
\begin{multline}\label{LIBOR-MP}
1+\delta L(\cdot,T_k)
 = \frac{M^{u_k}}{M^{\uk}} \in \mathcal{M}(\P_{T_{k+1}})\\
 \quad\text{ since }\quad
\frac{M^{u_{k}}}{M^{u_{k+1}}}
 \cdot \frac{\ud\P_{\tk[k+1]}}{\ud\P_{\tk[N]}}
 = \frac{M^{u_{k}}}{M^{u_{k+1}}}
  \cdot \frac{M^{u_{k+1}}}{M_0^{u_{k+1}}}
 \in \mathcal{M}(\P_{T_{N}}).
\end{multline}

Therefore, we have just described a broad framework for modeling \lib rates, in
which requirements (A1) and (A2) are satisfied. The next step is to show that
requirement (A3) is also satisfied. We will not pursue this in full generality,
instead we will consider a specific form for the variable $Y_T^u$, and thus for
the martingales $M^u$. In addition, the model is driven by an affine process,
and is henceforth called the \emph{affine LIBOR model}.

\subsection{Affine processes}
Let \prozess[X] be a stochastically continuous, time-homogeneous Markov process
with state space $D=\Rp^d$, starting from $x\in D$. The process $X$ is called
\emph{affine} if the moment generating function satisfies
\begin{align}\label{affine}
\E_x \big[\e^{\scal{u}{X_t}}\big]
 = \exp\big( \phi_t(u) + \scal{\psi_t(u)}{x} \big),
\end{align}
for some functions $\phi:[0,T]\times\I_T\to\R$ and $\psi:[0,T]\times\I_T\to\R^d$,
and all $(t,u,x)\in[0,T]\times\I_T\times D$, where
\begin{align}
\mathcal{I}_T
:= \set{u\in\R^d: \E_x\big[\e^{\scal{u}{X_T}}\big] < \infty,
        \,\,\text{for all}\; x \in D}.
\end{align}
We will assume in the sequel that $0\in\I_T^\circ$. The functions $\phi$ and
$\psi$ satisfy the semi-flow property
\begin{equation}\label{flow}
\begin{split}
\phi_{t+s}(u) &= \phi_{t}(u)+\phi_{s}(\psi_t(u))\\
\psi_{t+s}(u) &= \psi_{s}(\psi_{t}(u)),
\end{split}
\end{equation}
with initial condition
\begin{align}\label{phi-psi-0}
 \phi_0(u)=0
  \quad\text{ and }\quad
 \psi_0(u)=u,
\end{align}
for all $(t+s,u)\in[0,T]\times\I_T$. Equivalently, $\phi$ and $\psi$ satisfy
generalized Riccati differential equations. For comprehensive expositions of
affine processes we refer the reader to \shortciteN{DuffieFilipoviSchachermayer03}
and \citeN{KellerRessel08}.

\subsection{Affine \lib model}
In the \alm, the random variable $Y_T^u$ has the following form:
\begin{align}
Y_T^u = \e^{\scal{u}{X_T}},
\end{align}
where $u\in\Rp^d$ and $X_T$ is a random variable from an $\Rp^d$-valued affine
process $X$. Hence, $Y_T^u\ge1$, while the map $u\mapsto Y_T^u$ is obviously
increasing; note that inequalities involving vectors are understood componentwise.

Using the Markov property of affine processes, we can deduce that the martingales
$M^u$ have the form
\begin{align}\label{alm-Mu}
M_t^u
 &= \E\big[\e^{\scal{u}{X_T}}|\F_t\big] \nonumber\\
 &= \exp\big( \phi_{T-t}(u) + \scal{\psi_{T-t}(u)}{X_t} \big).
\end{align}
Therefore, \lib rates have the following evolution:
\begin{align}
1+\delta L(t,T_k)
 &= \frac{M_t^{u_k}}{M_t^{u_{k+1}}}
  = \exp\big(A_{k,t} + \scal{B_{k,t}}{X_t}\big),
\end{align}
where
\begin{equation}\label{Ak-Bk}
\begin{split}
 A_{k,t} &:= \phi_{T_N-t}(u_k) - \phi_{T_N-t}(u_{k+1})\\
 B_{k,t} &:= \psi_{T_N-t}(u_k) - \psi_{T_N-t}(u_{k+1}).
\end{split}
\end{equation}
Let us point that, under reasonable assumptions on the driving affine process,
we can prove that the \alm can fit \emph{any} term structure of 
initial \lib rates; cf. Proposition 6.1 in \krpt.

Now, regarding requirement (A3), let us turn our attention to the structure of
the driving process under the different forward measures. Using the connections
between forward measures \eqref{Pk-to-final}, the Markov property of affine
processes, and the flow equations \eqref{flow}, we can show that
\begin{align}
\E_{T_k}\Big[\e^{\langle v,X_r\rangle}\big|\F_s\Big]
 &= \E_{T_N}\Big[ \frac{M_r^{u_k}}{M_s^{u_k}} \e^{\langle v,X_r\rangle}\big|\F_s\Big] \\
 &= \exp\Big(\phi_{r-s}(\psi_{T_N-r}(u_{k})+ v) - \phi_{r-s}(\psi_{T_N-r}(u_k))\nonumber\\
 &\qquad
  + \scal{\psi_{r-s}(\psi_{T_N-r}(u_{k})+ v) - \psi_{r-s}(\psi_{T_N-r}(u_k))}{X_s}\Big);
  \nonumber
\end{align}
cf. \shortciteN[eq. (6.15)]{KellerResselPapapantoleonTeichmann09}. This means
that $X$ becomes a \emph{time-inhomogeneous affine} process under \emph{any}
forward measure. Note that the measure changes are again Esscher
transformations, similarly to \fpms. Consequently the \alm satisfies requirements
(A1), (A2) \emph{and} (A3).

The pricing of caplets and swaptions in the \alm using Fourier transform methods
is described in \krpt. In addition, closed-form valuation formulas -- in terms
of the $\chi^2$-distribution function -- are derived when the driving affine
process is the Cox--Ingersoll--\linebreak Ross (CIR) process.

\section{Extensions}
\label{extend}

The different approaches for modeling \lib rates have been extended in two
different directions: (i) to model simultaneously LIBOR rates for different
currencies and the corresponding foreign exchange rates, and (ii) to jointly
model default-free and defaultable \lib rates.

\subsection{Multiple currencies}
The log-normal \lmm has been extended to a multi-currency setting by \citeN{Schloegl02}
and by \citeANP{Mikkelsen02} \citeyear{Mikkelsen02}. The \lev \lib model and the \lev \fpm have
been extended to model multiple currencies and foreign exchange rates by
\citeN{EberleinKoval06}. A multi-factor approach to multiple currency \lib models
has been presented in \shortciteN{BennerZyapkovJortzik09}. \mfms have been extended
to the multi-currency setting by \citeANP{FriesRott04} \citeyear{FriesRott04} and
\citeN{FriesEcksteadt08}.

\subsection{Default risk}
The log-normal \lmm has been first extended to model default risk by
\citeN{LotzSchloegl00}, who used a deterministic hazard rate to model the time
of default. \shortciteN{EberleinKlugeSchoenbucher06}, \linebreak borrowing also ideas from
\citeN{Schoenbucher00}, constructed a model for default-free and defaultable
rates where they use time-inhomogeneous \lev processes as the driving motion and
the ``Cox construction'' to model the time of default (cf. e.g.
\citeN{BieleckiRutkowski02} for the Cox construction). This has been extended
to a model where defaultable bonds can have rating migrations by \citeN{Grbac09}.

\bibliographystyle{chicago}
\bibliography{references}

\end{document}